\documentclass[11pt]{article}
\pdfoutput=1

\usepackage{jheppub}
\usepackage{graphicx}
\usepackage{amsmath}
\usepackage{amssymb}
\usepackage{amsfonts}
\usepackage{bm}
\usepackage{mathtools}
\usepackage{braket}
\usepackage{hyperref}
\usepackage{wrapfig}

\hypersetup{colorlinks=true,urlcolor=[rgb]{0,0,0.5},citecolor=[rgb]{0.5,0,0},linkcolor=[rgb]{0,0,0.4}}

\def\Tr{{\rm Tr}}
\def\op{{\cal O}}

\def\vev#1{\langle{#1}\rangle}

\def\where{\quad {\rm where} \quad}
\def\for{\quad {\rm for} \quad}
\def\and{\quad {\rm and} \quad}

\def\nn{\nonumber\\}
\def\ra{\rightarrow}
\def\ie{{\rm i.e.\ }}
\def\eg{{\rm e.g.\ }}

\def\CC{{\cal C}}

\def\CE{{\cal E}}
\def\CF{{\cal F}}
\def\CH{{\cal H}}
\def\CJ{{\cal J}}

\def\CN{{\cal N}}
\def\CR{{\cal R}}

\title{Chaos and random matrices in supersymmetric SYK}
\subheader{{\rm \hfill CALT-TH-2017-056}}

\author[a]{Nicholas Hunter-Jones}
\author[b]{and Junyu Liu}
\affiliation[a]{Institute for Quantum Information and Matter,\\ California Institute of Technology, Pasadena, California 91125, USA}
\affiliation[b]{Walter Burke Institute for Theoretical Physics,\\ California Institute of Technology, Pasadena, California 91125, USA}

\emailAdd{nickrhj@caltech.edu}
\emailAdd{jliu2@caltech.edu}

\abstract{
We use random matrix theory to explore late-time chaos in supersymmetric quantum mechanical systems. Motivated by the recent study of supersymmetric SYK models and their random matrix classification, we consider the Wishart-Laguerre unitary ensemble and compute the spectral form factors and frame potentials to quantify chaos and randomness. Compared to the Gaussian ensembles, we observe the absence of a dip regime in the form factor and a slower approach to Haar-random dynamics. We find agreement between our random matrix analysis and predictions from the supersymmetric SYK model, and discuss the implications for supersymmetric chaotic systems.
}

\begin{document}

\maketitle
\flushbottom

\section{Introduction}
A recent surge of interest in quantum chaos has revolved around a strongly-interacting quantum system called the Sachdev-Ye-Kitaev (SYK) model \cite{Kitaev15,SachdevYe}. This model of $N$ all-to-all randomly interacting Majorana fermions is solvable at strong-coupling and appears to be in the same universality class as black holes, exhibiting an emergent reparametrization invariance and an extensive ground-state entropy. More compellingly, the out-of-time order correlation function (OTOC) of the theory \cite{Kitaev15,MS_SYK} saturates a universal bound on chaotic growth \cite{MSSbound}, a seemingly unique feature of gravity \cite{SSbutterfly,SSstringy} and conformal field theories with a holographic dual \cite{ChaosCFT}. The low-energy description of the theory in terms of a Schwarzian effective action also encapsulates dilaton gravity in AdS$_2$ \cite{Jensen16,SYKbulk}. This model should be seen as a valuable resource for understanding both black holes and quantum chaos. 

There have already been a myriad of generalizations of the SYK model, including an extension by Fu, Gaiotto, Maldacena, and Sachdev, to a supersymmetric model of strongly interacting Majoranas \cite{SUSY_SYK}, which has been further explored in \cite{Peng:2016mxj, Murugan:2017eto, Yoon:2017gut, Peng:2017spg, Kanazawa:2017dpd}. The supersymmetric version of the model also displays many of the same holographic properties. Notably, at strong-coupling the theory has an emergent superconformal symmetry which renders it solvable and allows one to compute correlation functions. At low-energies the symmetry is broken, giving a Schwarzian-like effective action which mimics supergravity in AdS$_2$ \cite{SchwSugra}. Like its non-supersymmetric counterpart, the model has random matrix universality in its spectral statistics \cite{You16,SUSY_RMT} and appears to exhibit thermalization in its eigenstates \cite{Sonner17,SYK_ETH}, both hinting at underlying chaotic dynamics. 

Although we lack a precise definition of quantum chaos, there are still universal features one expects of quantum chaotic systems: most notably, having the spectral statistics of a random matrix \cite{BGSchaos}. Information scrambling \cite{HaydenPreskill,SekinoSusskind} and chaotic correlation functions \cite{SSbutterfly} have also been extolled as symptoms of chaos. Ideas from quantum information have helped make these notions more precise, quantifying how scrambling \cite{ChaosChannels} and randomness \cite{ChaosDesign} are encoded in OTOCs. Similarly, \cite{ChaosRMT} explored the connection to random matrix dynamics, quantifying randomness and scrambling in the time evolution by random matrix Hamiltonians and computing a quantity called the frame potential. The onset of random matrix behavior can also be seen in the spectral form factor, which has been studied in the SYK model \cite{BHRMT16}.

Motivated by this, we may ask the question: what are the universal features of supersymmetric SYK models, or more generally, of all supersymmetric quantum chaotic systems? And how do we quantify them from an information-theoretic standpoint?

To address this, we consider the Wishart-Laguerre ensembles, also termed random covariance matrices \cite{TaoRMT}, which appeared in the random matrix classification of the supersymmetric SYK models \cite{SUSY_RMT}. Recall that the Hamiltonian in supersymmetric quantum mechanics is constructed as the square of a supercharge. Loosely speaking, the intution is that this random matrix ensemble arises from squaring the Gaussian random matrices, just as we might think of a chaotic supersymmetric system defined by a disordered supercharge. In this paper we consider the simplest Wishart-Laguerre ensemble,\footnote{Interestingly, Wishart ensembles have appeared in studying the reduced density matrix in systems evolved with random matrix Hamiltonians \cite{VZsub}. Wishart ensembles have also appeared in random matrix contructions of supergravity to explore the space of AdS vacua \cite{Marsh12}.} the Wishart-Laguerre unitary ensemble (LUE), corresponding to supersymmetric quantum systems without additional discrete symmetries. In the following, we will quantitatively derive predictions for the spectral form factors, frame potential, and the out-of-time-ordered correlators, where a central distinction from the non-supersymmetric models arises in the spectral 1-point functions, which modifies the early time decay of the spectral form factor. A slower decay in the LUE frame potential indicates less efficient information scrambling and the failure of the ensemble to become Haar-random. Our predictions for the LUE match those from the 1-loop partition function of the supersymmetric SYK model.

The paper is organized as follows: In Section \ref{sec:setup}, we review the supersymmetric model and spectral form factor, discussing its universal features and behavior in SYK models. In Section \ref{sec:SFF}, we review the basic tools in random matrix theory and then compute spectral form factors for the Wishart-Laguerre ensemble. In Section \ref{sec:LUEch}, we explore chaos in this random matrix ensemble by computing the frame potentials and correlation functions, and comment on its complexity growth. In Section \ref{sec:SYKch}, we discuss chaos in supersymmetric SYK and compare with the random matrix predictions, concluding in Section \ref{sec:con}. In Appendix \ref{app:num} we present some numerical checks of our expressions.

\section{Setup and overview}
\label{sec:setup}

\subsection{Supersymmetric SYK model}
We first briefly review the supersymmetric extension of the SYK model. For an in-depth discussion of the original model, see \cite{MS_SYK}. Consider $N$ all-to-all interacting Majorana fermions $\psi_i$ with random couplings, which anticommute as $\{\psi_i, \psi_j \} = \delta_{ij}$. The $(2q-2)$-point $\CN=1$ supersymmetric model is constructed from the supercharge $Q$, a $q$-body Majorana interaction with odd $q$. The Hamiltonian is then given by the square of the supercharge as
\begin{equation}
H = Q^2\,, \where Q = i^{(q-1)/2} \sum_{i_1<\ldots<i_q} C_{i_1\ldots i_q} \psi_{i_1}\ldots \psi_{i_q}\,,
\end{equation}
with Gaussian random couplings $C_{i_1\ldots i_q }$ of mean and variance
\begin{equation}
\big\langle C_{i_1\ldots i_q} \big\rangle = 0\,, \qquad \big\langle C_{i_1\ldots i_q}^2 \big\rangle = \frac{J^2 (q-1)!}{N^{q-1}}\,,
\end{equation}
and where $J$ is a positive constant. We also define $\CJ$ as $J^2 = 2^{q-1} \CJ^2 /q$, with a slightly more convenient scaling in $q$. 

In the large $N$ limit, this model shares many of the same appealing holographic features as the SYK model, such as chaotic correlation functions, a zero-temperature entropy, and an emergent superconformal symmetry which is broken at low-energies, admitting a Schwarzian-like desciption \cite{SUSY_SYK}. We can compute the free energy at large $N$ by evaluating at the saddle point, and at low temperatures find
\begin{equation}
\log Z = -\beta E_0 + N s_0 + \frac{cN}{2\beta} + \ldots\,,
\label{eq:part}
\end{equation}
where $s_0$ is the zero-temperature entropy density and $c$ is the specific heat. In the supersymmetric theory we have $c = \alpha \pi^2/ \CJ $ with a constant $\alpha$, which becomes $ c = \pi^2 / 4 q^2 \CJ $ in the large $q$ limit. The ground-state entropy density is computed to be $s_0 = \frac{1}{2} \log (2 \cos \frac{\pi}{2 q})$ and the ground state energy $E_0$ can be subtracted off. 

The SYK model with $N$ Majoranas enjoys a random matrix classification, where the symmetry class of the theory is dictated by a particle-hole symmetry \cite{You16,BHRMT16}. Depending on $N$, the spectrum will display level statistics of one of the three Gaussian ensembles: GUE, GOE, or GSE. %, realizing the three symmetry classes of Dyson \cite{DysonSym}. 
For the supersymmetric extension of SYK, we can similarly classify the random matrix behavior for a given number of Majoranas $N$, going beyond Dyson's classification to the extended 10-fold symmetry classification of Altland-Zirnbauer \cite{AZsym}. Understanding how anti-unitary symmetries act on the supercharge $Q$, we can identify the appropriate symmetry class \cite{SUSY_RMT}. The Hamiltonian, given as the square of the supercharge, then has random matrix description in terms of the Wishart-Laguerre ensembles. The level statistics are still those of the Gaussian ensembles, but the spectral correlations are different. Roughly, we can think of the supersymmetric SYK behaving like the square of Gaussian random matrices, which are the Wishart ensembles. For more details, see \cite{SUSY_RMT} as well as an extension of the classification to the $\CN=2$ supersymmetric models \cite{Kanazawa:2017dpd}.

Speaking generally, there a number of reasons one might wish to consider supersymmetric generalizations of SYK. For instance, much is understood about the low-energy physics in nearly AdS$_2$ spacetimes purportedly dual to the low-energy dynamics in SYK, but the exact holographic dual of the theory is not known. As many of the best understood examples of AdS/CFT are supersymmetric, one might hope that this particular construction might provide guidance on the correct UV completion of the SYK model. Less ambitiously, considering the supersymmetric models might be useful in contructing higher dimension analogs \cite{Murugan:2017eto}.

\subsection{Spectral form factor}
Quantum chaotic systems are often defined to have the spectral statistics of a random matrix. An object familiar in random matrix theory which exhibits these universal properties is the spectral form factor. We will introduce this object more precisely in our review of random matrix theory in Sec.~\ref{sec:RMT}, but the 2-point spectral form factor $\CR_2(t,\beta)$ can be given simply in terms of the analytically continued partition function
\begin{equation}
\CR_2(t,\beta) \equiv \big\langle Z(\beta, t)Z^*(\beta, t)\big\rangle\,, \where Z(\beta, t) \equiv \Tr \big( e^{-\beta H - it H}\big)\,,
\end{equation}
and where the average $\langle \,\cdot\, \rangle$ is taken over an ensemble of Hamiltonians (\eg SYK, or some disordered spin system, or a random matrix ensemble). This object was discussed more recently in \citep{BHRMT16}, where they studied the form factor in SYK and found that the theory revealed random matrix behavior at late times. From the bulk point of view, one motivation for studying this object was a simple version of black hole information loss \cite{MaldacenaEternal}: 2-point functions appear to decay exponentially in terms of local bulk variables, whereas a discrete spectrum implies a finite late-time value. The same inconsistency is apparent in the spectral form factor. 

\begin{figure}[htb!]
\centering
\includegraphics[width=0.95\linewidth]{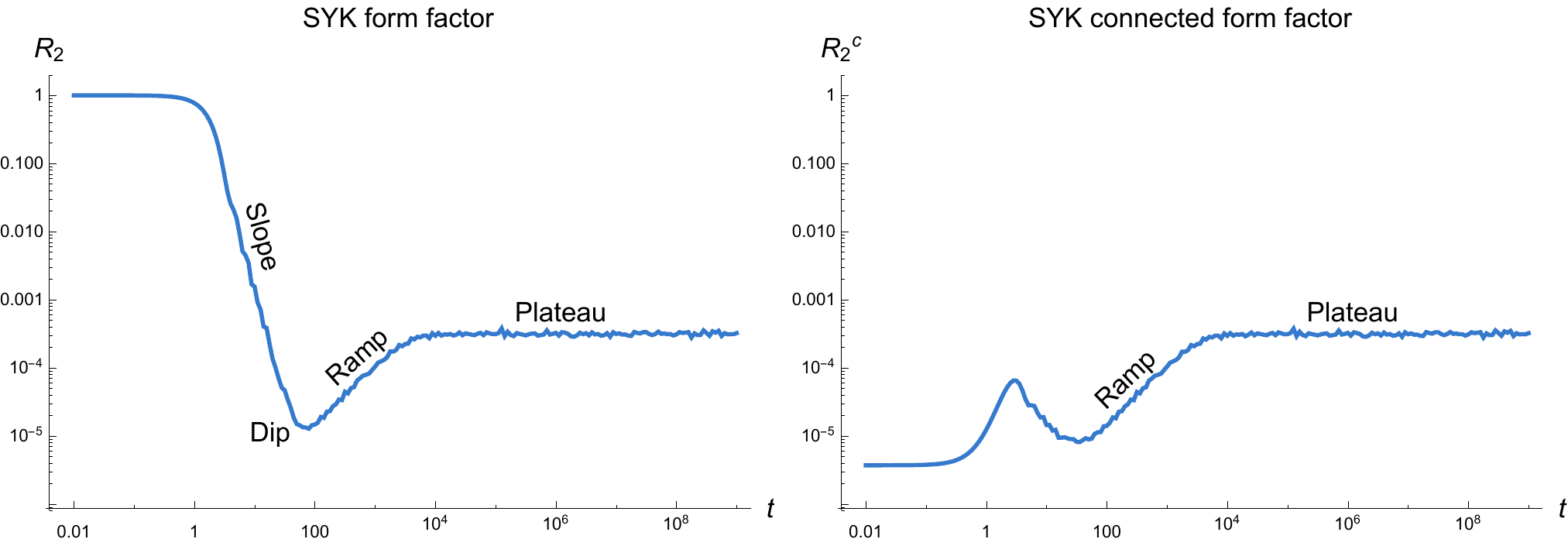}
\caption{The $2$-point spectral form factor and its connected component for SYK with $N=24$ Majoranas at inverse temperature $\beta =1$, computed for 800 realizations of disorder. We observe the slope, dip, ramp, and plateau behaviors. 
}
\label{fig:SYKff}
\end{figure}

Some characteristic features of the time-evolved form factor $\CR_2(t)$, exhibited in both the SYK model and in random matrix theories, are: an early time decay from an initial value called the slope, a crossover at intermediate times called the dip, a steady linear rise called the ramp, and a late-time floor called the plateau. In Fig.~\ref{fig:SYKff} we observe these features in SYK. While the early time decay depends on the specific system, the ramp and plateau should be universal features of quantum chaotic systems. The ramp is characteristic of spectral rigidity: the long-range logarithmic repulsion of eigenvalues. The anticorrleation of eigenvalues causes the linear increase in the form factor. At late times, or at energy scales smaller than the mean spacing, the form factor reaches a plateau as degeneracies are rare and neighboring eigenvalues repel in chaotic systems. 

\subsection*{SYK form factor and GUE}
Recently, \cite{BHRMT16} studied the form factor in SYK and found agreement with random matrix theory, showing analytically and numerically the aspects of the dip, ramp, and plateau of SYK agree with those of the Gaussian unitary ensemble (GUE), an ensemble of $L\times L$ random Hermitian matrices. We will avoid explicitly introducing and defining the original Majorana, instead simply mentioning a few details to better frame the discussion of the model's supersymmetric extension. 

The emergent reparamentrization invariance of SYK at strongly-coupled is broken spontaneously and explicitly at low-energies, yielding an effective description in terms of the Schwarzian derivative \cite{Kitaev15,MS_SYK}. The 1-loop partition function of the Schwarzian theory $Z^\text{Sch}_\text{1-loop} \sim e^{cN/2\beta}/\beta^{3/2}$, can be analytically continued to $\beta + i t$ to study the form factor of SYK. At early times, $\CR_2(t,\beta)$ is dominated by the disconnected piece which gives a $1/t^3$ power law decay, normalized by its initial value we have
\begin{equation}
\frac{\langle Z(\beta,t)Z^*(\beta,t) \rangle}{\vev{Z(\beta)}^2} \simeq \frac{\beta^3 e^{-c N/\beta}}{t^3}\,,
\end{equation}
for times greater than $t \gtrsim \sqrt{N} $ when the time dependence in the exponent disappears and where $c$ is the specific heat of the theory. To isolate this contribution, \cite{BHRMT16} considered a special limit (a `triple scaled' limit) where only the Schwarzian contributes. Moreover, \cite{FermiLoc} showed that the Schwarzian theory is 1-loop exact and recieves no higher-order corrections, indicating that the power-law decay predicted by the Schwarzian should dominate the disconnected form factor for long times.\footnote{For more on solving the Schwarzian theory, see \cite{Altland16,SchwBoot}.} This power law decay is simply the Laplace transform of the statement that the spectrum has a square-root edge\footnote{As discussed in \cite{MS_SYK}. The spectral density of SYK has been further studied in \cite{BHRMT16,Garcia16,Garcia17}.}
\begin{equation}
\rho(E) \sim \sinh \sqrt{2c E N}\,.
\end{equation}
Knowing the free energy in the large $N$ limit, we can also show that the form factor of SYK transitions to a ramp at a dip time $t_d \sim e^{Ns_0/2}$, growing linearly until a plateau time of $t_p \sim e^{Ns_0 + cN/2\beta}$, where $s_0$ is the zero-temperature entropy density. 

Many of these features of the SYK form factor agree with the universal predictions from GUE. The form factor for GUE has been studied extensively in the random matrix literature \cite{MehtaRMT,BrezinHikami1,BrezinRMT} and references therein, and revisited more recently in the context of SYK and black holes in \cite{BHRMT16,delCampoScrambFF,ChaosRMT}. Simply stating the results, the early-time decay of the GUE form factor transitions to a linear ramp at a dip time of $t_d \sim \sqrt{L}$, growing linearly until the plateau time $t_p \sim L$. We note that around the plateau time the ramp is not quite linear as nonperturbative effects become important as we transition to the plateau \cite{AAinst}. The non-universal early time decay also has the same power law $1/t^3$, due to the fact the Wigner semicircle law for Gaussian random matrices $\rho(\lambda)= \frac{1}{2\pi}\sqrt{4-\lambda^2}$, also exhibits a square-root edge.

\subsection*{Supersymmetric SYK form factor}
From the large $N$ partition function of the supersymmetric theory, we can also make predictions as to the behavior of the spectral form factor. We will present a more explicit treatment of this in Sec.~\ref{sec:SYKch}. At low-energies, the fluctuations around the large $N$ saddle point of the supersymmetric theory break superconformal symmetry; the action for these reparametrizations is a super-Schwarzian \cite{SUSY_SYK}, where the action integrates over $\tau$ and a superspace coordinate $\theta$ and the super-Schwazian acts just like the standard Schwarzian derivative except as a super-derivative, respecting a similar chain rule. The action gives a 1-loop partition function
\begin{equation}
Z^\text{sSch}_\text{1-loop}(\beta) \sim \frac{1}{\sqrt{\beta \CJ}} e^{Ns_0 + cN/2\beta}\,,
\end{equation}
which differs in the 1-loop determinant from the SYK model. The super-Schwarzian theory is also 1-loop exact \cite{FermiLoc}, ensuring its validity away from very early times. Analytically continuing the partition function $\beta \ra \beta+it$, disconnected piece of the form factor which dominates at early times, is
\begin{equation}
\frac{\vev{Z(\beta,t)Z^*(\beta,t)}}{\vev{Z(\beta)}^2} \simeq \frac{\beta e^{-cN/\beta}}{t}\,,
\end{equation}
exhibiting a $1/t$ decay in the slope, slower than the decay in SYK. This can also be understood as the contribution from the edge of the spectrum, where the Laplace transform of the 1-loop partition function gives
\begin{equation}
\rho(E) \sim \frac{1}{\sqrt{\CJ E}} \cosh \big( \sqrt{2 c N  E} \big) \,,
\end{equation}
observing a square-root growth at the edges of the spectrum. 

\begin{figure}[htb!]
\centering
\includegraphics[width=0.95\linewidth]{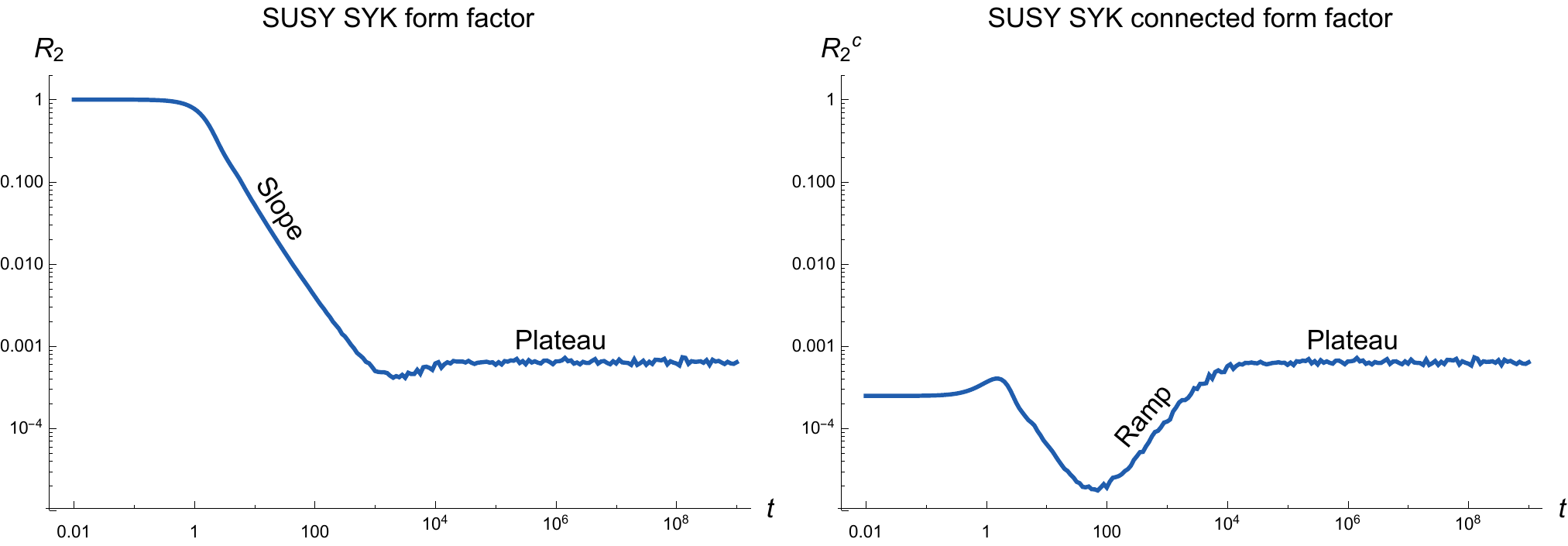}
\caption{The $2$-point spectral form factor and its connected piece for the supersymmetric SYK model with $N=24$ Majoranas at inverse temperature $\beta =1$, computed for 800 realizations of disorder. We observe the slope and plateau behaviors, while the ramp is obscured by the slow early-time decay of the 1-point function.
}
\label{fig:susySYKff}
\end{figure}

As we discuss later, computing the ramp function for supersymmetric SYK, we find the ramp and slope intersect at a dip time $t_d \sim e^{Ns_0}$, which is the same time scale as the ramp's transition to the plateau $t_p \sim e^{Ns_0}$. The slow decay at early times means that the slope transitions to ramp behavior at the same time-scale as the plateau time, \ie the ramp is hidden beneath the slope. We plot the 2-point form factor for the model in Fig.~\ref{fig:susySYKff}. Subtracting the disconnected contribution reveals the ramp in the connected form factor, also plotted. The lack of a dip in the supersymmetric model will have implications for our discussion of the frame potential and randomness.

\subsection*{Notation}
A brief comment on notation. In recent work studying the spectral form factor, the normalized 2-point form factor is often denoted as $g(t,\beta)$, and its connected component as $g_c(t,\beta)$:
\begin{equation}
g(t,\beta) \equiv \frac{\vev{Z(\beta,t)Z^*(\beta,t)}}{\vev{Z(\beta)}^2} \and g_c(t,\beta) \equiv g(t,\beta) - \frac{\vev{Z(\beta,t)}\vev{Z^*(\beta,t)}}{\vev{Z(\beta)}^2}\,.
\end{equation}
While in \cite{ChaosRMT}, we denoted the 2-point form factor as $\CR_2(t,\beta)$, and more generally the $2k$-th form factor as $\CR_{2k}(t,\beta)$. Just to be clear
\begin{equation}
g(t,\beta) = \frac{\CR_2(t,\beta)}{\vev{Z(\beta)}^2}\,, \quad\text{or at } \beta=0: \quad g(t,0) = \frac{\CR_2(t)}{L^2}\,.
\end{equation}
For us, working directly with the numerator turns out to be more convenient when discussing the frame potential and correlation functions, and avoids subtleties regarding the appropriate or tractable normalization, \ie `quenched' vs `annealed'. 

\section{Form factors for Wishart matrices}
\label{sec:SFF}
\subsection{Basic setup in random matrix theory}
\label{sec:RMT}
In this paper, we consider the Wishart-Laguerre Unitary Ensemble (LUE), an ensemble of $L\times L$ random matrices which can be generated as $H^\dagger H$, where $H$ is a complex Gaussian random matrix with normally distributed complex entries drawn with mean 0 and variance $\sigma^2 = 1/L$. This is the `physics normalization', where the spectrum does not scale with system size.\footnote{Note that it is common in the random matrix literature to instead work with unit variance $\sigma^2=1$.} The joint probability distribution of LUE eigenvalues is given by
\begin{align}
P(\lambda )d\lambda = C \left| \Delta (\lambda ) \right|^2 \prod_{k=1}^L e^{-\frac{L}{2} \lambda_k} d\lambda_k\,,
\end{align}
where $\Delta (\lambda )$ is the Vandermonde determinant and the constant factor is defined such that the distribution integrates to unity. One can think of LUE matrices as square of a Gaussian random matrix. More generally, we could define $L\times L$ Wishart matrices generated by $L'\times L$ Gaussian matrices, where $L'\geq L$, which gives a slightly more general eigenvalue distribution. But given the supersymmetric Hamiltonians we consider defined as the square of the supercharge, we just consider Wishart matrices generated by square matrices with $L=L'$. We average over the random matrix ensemble as
\begin{equation}
\vev{\op} \equiv \int D\lambda \,\op \where \int D\lambda = C \int \prod_k d\lambda_k |\Delta(\lambda)|^2 e^{-\frac{L}{2} \sum_k \lambda_k}\,.
\end{equation}
The spectral density is given by integrating the joint probability $P(\lambda)$ over $L-1$ variables
\begin{align}
\rho (\lambda )=\int d \lambda_1 d\lambda_2\ldots d\lambda_{L-1}\, P(\lambda_1, \lambda_2, \ldots, \lambda_{L-1},\lambda)\,.
\end{align}
More generally, we can define the $k$-point spectral correlation function by integrating over all but $k$ arguments
\begin{align}
\rho^{(k)} (\lambda_1, \lambda_2, \ldots, \lambda_k) = \int d\lambda_{k+1} d\lambda_{k+2} \ldots d\lambda_L\, P(\lambda_1, \lambda_2, \ldots, \lambda_k, \lambda_{k+1}, \ldots, \lambda_L) \,.
\end{align}
%It will also be convenient to define a rescaled form of the $k$-point correlation functions\footnote{In the RMT literature, $R^{(k)}$ is sometimes the object referred to as the $k$-point correlator.}
%\begin{align}
%R^{(k)} (\lambda_1, \lambda_2, \ldots, \lambda_k) = \frac{L!}{(L-k)!} \rho^{(k)} (\lambda_1, \lambda_2, \ldots, \lambda_k)\,,
%\end{align}
%with combinatorial factors to account for the freedom to permute the arguments of $\rho^{(k)}(\lambda_1,\ldots,\lambda_k)$. 
Recall that for the Gaussian ensembles, we may take the large $L$ limit famously recover Wigner's semicircle law for the distribution of eigenvalues. Instead in the LUE, we take the large $L$ limit and find \cite{MarPas}
\begin{align}
\rho(\lambda)=\frac{1}{2\pi \lambda}\sqrt{\lambda(4-\lambda)}\,,
\end{align}
which is referred to as the Mar{\v c}enko-Pastur distribution.

Just as in the GUE, the LUE is a determinantal point process, which means the $k$-point spectral correlators are given by a kernel $K$ as
\begin{equation}
\rho^{(k)} (\lambda_1,\ldots, \lambda_k) = \frac{(L-k)!}{L!} \det \big( K(\lambda_i, \lambda_j) \big)_{i,j=1}^{k}\,.
\label{eq:det}
\end{equation}
Demonstrating the universality of Dyson's sine kernel \cite{DysonSK}, the Wishart ensemble has sine kernel statistics in the large $L$ limit \cite{Nagao91,TaoRMT}, meaning
\begin{equation}
K(\lambda_i, \lambda_j) = 
\left\{\begin{array}{*{35}{l}} \dfrac{\sin \big(L\rho (u)\pi (\lambda_i - \lambda_j)\big)}{\pi (\lambda_i-\lambda_j) } & \for i\ne j  \vspace*{12pt}\\
\dfrac{L}{2\pi \lambda_i} \sqrt{ \lambda_i (4-\lambda_i) }\  & \for i=j \,, \\
\end{array} \right.
\end{equation}
where $u$ is an arbitrary constant valued in $[0,4]$. We will fix the value of $u$ numerically.\footnote{The analogous constant in considering the GUE would be fixed to $u=0$, given the symmetry of the spectrum. However, for the LUE $u=0$ it is divergent. The value of $u$ specifies the center of the two eigenvalues $\lambda_i$ and $\lambda_j$.}

The spectral form factor, defined as the Fourier transform of the spectral correlation functions, is a standard quantity to consider in random matrix theory; see \cite{MehtaRMT} for an overview. We define the 2-point spectral form factor in terms of the analytically continued partition function $Z(\beta, t)$ as\footnote{This is slightly different than the standard presentation in the RMT literature, where the form factor is usually given as the Fourier transform of a connected form factor, called the cluster function. Here we work with both connected and disconnected pieces.}
\begin{equation}
\CR_2(t,\beta) \equiv \left\langle Z(\beta ,t){{Z}^{*}}(\beta ,t) \right\rangle = \int D\lambda \sum_{i,j} e^{i(\lambda_i-\lambda_j)t} e^{-\beta(\lambda_i+\lambda_j)}\,,
\end{equation}
where the continued partition function $Z(\beta,t)$ is
\begin{equation}
Z(\beta ,t)=\text{Tr}\left( {{e}^{-\beta H-iHt}} \right)\,.
\end{equation}
More generally, we consider $k$-point spectral form factors which we define as
\begin{align}
\CR_{2k}(t,\beta) &\equiv \left\langle \left( Z(\beta, t) Z^* (\beta ,t) \right)^{2k} \right\rangle\\
&=\int D\lambda \sum_{i,j} e^{i(\lambda_{i_1}+\ldots+\lambda_{i_k}-\lambda_{j_1}-\ldots \lambda_{j_k} )t} e^{-\beta (\lambda_{i_1}+\ldots+\lambda_{i_k}+\lambda_{j_1}+\ldots+\lambda_{j_k})}\,.
\end{align}
In the following subsections, we will compute the LUE spectral form factors and compare analytical results with numerical observations. 

At large $L$, we compute the spectral form factors by Fourier transforming the determinant of kernels in Eq.~\eqref{eq:det}. We integrate the products of $K$ as \cite{MehtaRMT}
\begin{align}
&\int \left( \prod_{j=1}^n d\lambda_j\, e^{i k_j \lambda_j} \right) \, K(\lambda_1, \lambda_2) K(\lambda_2,\lambda_3) \ldots K(\lambda_{n-1}, \lambda_n) K(\lambda_n,\lambda_1) \nn
&\qquad = \alpha_L \int d\lambda\, e^{i \sum_{j=1}^n k_j \lambda} \int dk\, g(k) g\Big(k+\frac{k_1}{2\pi \alpha_L} \Big) g\Big(k+\frac{k_2}{2\pi \alpha_L} \Big) \ldots g\Big(k+\frac{k_{n-1}}{2\pi \alpha_L} \Big)\,,
\label{eq:int}
\end{align}
where the Fourier transform of the sine kernel is
\begin{equation}
g(k) \equiv \int dr\, e^{2\pi i k r}\, \frac{\sin (\pi r)}{\pi r} = \bigg\lbrace \begin{array}{c}1 ~{\rm for}~ |k|<1/2\\ 0 ~{\rm for}~ |k|>1/2 \end{array}\,,
\end{equation}
and where $\alpha_L \equiv L \rho(u)$. The integral over the sine kernel is unbounded and can be treated by imposing a cutoff. We use the \emph{box approximation} \cite{ChaosRMT}
\begin{equation}
\alpha_L \int d\lambda\, e^{i \sum_{j=1}^n k_j \lambda} \to \alpha_L \int_{-L/2\alpha_L}^{L/2\alpha_L} d\lambda \, e^{i \sum_{j=1}^n k_j \lambda} = L \frac{\sin \big(\sum_{j=1}^n k_j /2\rho (u) \big)}{\sum_{j=1}^n k_j /2\rho (u)}\,,
\end{equation}
fixed such that Eq.~\eqref{eq:int} over the truncated range with $k_i=0$ integrates to $L$. This will be helpful in computing the higher-point spectral form factors, for instance, $\mathcal{R}_4$. 

It will also be convenient to define the following functions which will appear in computing the LUE form factors
\begin{align}
&r_1(t) \equiv e^{2it} \big( J_0(2t) - i J_1(2t) \big) \nn
&r_2(t) \equiv \left\{
\begin{array}{cl} 1-\dfrac{t}{2\pi L\rho (u)} & \for 0<t<2\pi L\rho (u)  \\
0  & \for t>2\pi L\rho (u) \end{array} 
\right. \nn
 &r_3(t) \equiv \frac{\sin \big(t/2\rho (u)\big)}{t/2\rho (u)}\,.
\label{eq:rfuncs}
\end{align}

\subsection{Two-point form factor at infinite temperature}
Let us start with the simplest case, the two point spectral form factor at infinite temperature $\beta=0$. Pulling out coincident eigenvalues, we have
\begin{equation}
\CR_2 (t) = \int D\lambda \sum_{i,j} e^{i(\lambda_i - \lambda_j)t} = L + L(L-1)\int d\lambda_1 d\lambda_2\, \rho^{(2)}(\lambda_1, \lambda_2) e^{i(\lambda_1-\lambda_2)t}\,.
\end{equation}
The determinant of kernels in Eq.~\eqref{eq:det} gives a squared 1-point function and 2-point function contribution. Using the integration formula in Eq.~\eqref{eq:int}, we obtain
\begin{equation}
\CR_2 (t) = L+L^2 |r_1(t)|^2-Lr_2(t)
\label{eq:LUER2}
\end{equation}
in terms of the functions defined above, and where
\begin{align}
|r_1(t)|^2=J_{0}^{2}(2t)+J_{1}^{2}(2t)\,.
\end{align}

\begin{figure}[htb!]
\centering
\includegraphics[width=0.95\linewidth]{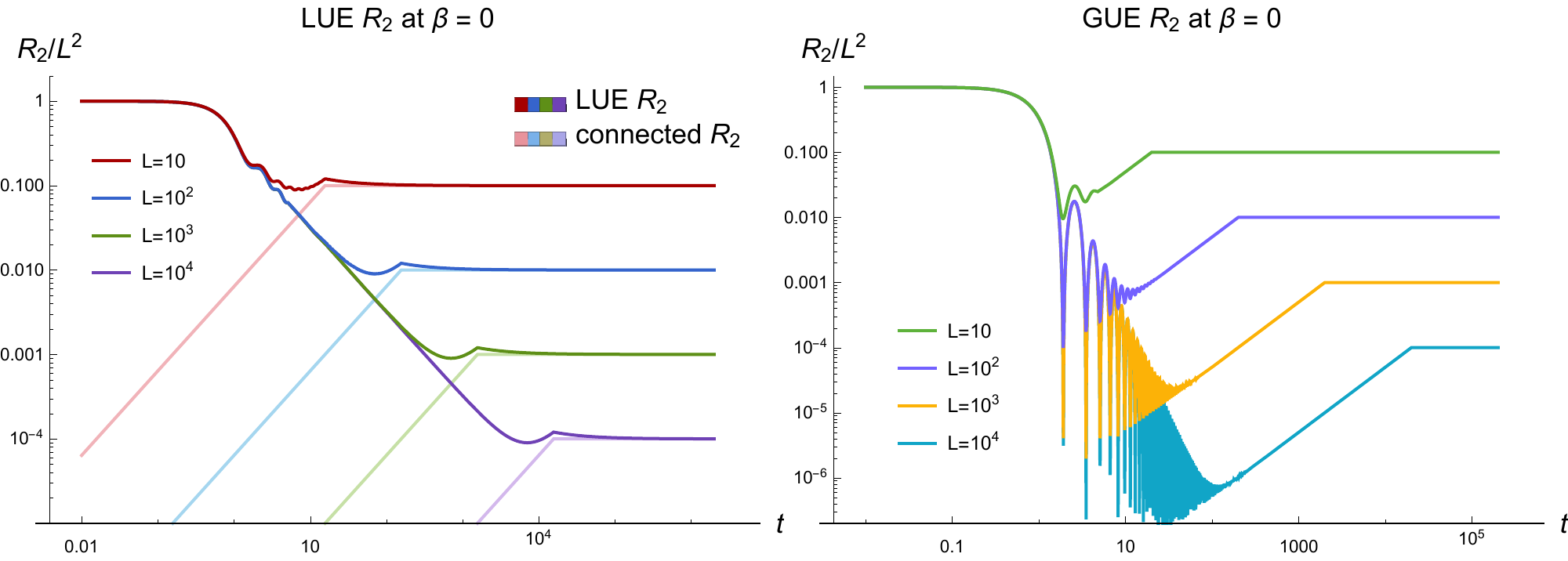}
\caption{On the left: the $2$-point spectral form factor and its connected component for the LUE at infinite temperature, as given in Eq.~\eqref{eq:LUER2}, plotted for different values of $L$ and normalized by the initial value $L^2$. We observe the slow $1/t$ decay down to the plateau value, hiding the linear ramp in the connected piece. On the right: the $2$-point spectral form factor for the GUE at infinite temperature, with a faster early-time decay exposing the ramp. 
}
\label{fig:R2plots}
\end{figure}

In Fig.~\ref{fig:R2plots}, we plot the infinite temperature LUE 2-point form factor as derived in Eq.~\eqref{eq:LUER2} along side the GUE form factor (see \cite{ChaosRMT}). Note that unlike in the GUE case there is no dip or ramp. The lack of an intermediate time scale at which the initial slope decay transitions at the dip to a linear growth to a plateau, is due to the slow decay of the 1-point functions which gives the slope.

Subtracting off the contribution from the 1-point functions defines the connected piece of the 2-point form factor
\begin{equation}
\CR^c_2 (t) \equiv \big\langle |Z(\beta ,t)|^2 \big\rangle - \big\langle Z(\beta ,t)\big\rangle^2 = L - Lr_2(t)\,,
\end{equation}
which exposes the linear growth before the plateau. The connected components are also plotted in Fig.~\ref{fig:R2plots}. 

The transition point in the function of $r_2$ is defined as the plateau time $t_p = 2\pi \alpha_L$, where $\alpha_L = L\rho(u)$. The value of $2\pi \alpha_L$ is not straightforwardly fixed given the unbounded support when integrating over kernels. The constant also determines the linear slope of the ramp function $r_2$ prior to the plateau. As we discuss in App.~\ref{app:num}, the constant $u$ is fixed by numerically fitting to the ramp. We find a plateau time of $t_p \sim \pi L /2$ for the LUE 2-point form factor.

Using the asymptotic form of the Bessel function,
\begin{equation}
J_k (z) \sim \sqrt{\frac{2}{\pi z}}\cos \Big(z-\frac{k \pi }{2}-\frac{\pi }{4} \Big)\,,
\end{equation}
we conclude that the disconnected piece decays at early times (for $t$ much smaller than $L$ but larger than $\op (1)$) as
\begin{equation}
r_1(t) r_1^*(t) = J_0^2(2t) + J_1^2(2t) \sim \frac{1}{\pi t} \big(\cos^2(2t-\pi/4) + \sin^2(2t-\pi/4)\big) = \frac{1}{\pi t}\,.
\end{equation}
This $\op(1/t)$ decay of the LUE form factor is to be contrasted with the slower $\mathcal{O}(1/t^3)$ decay in both the GUE and the SYK model \cite{ChaosRMT,BHRMT16}. However, the connected piece, dominated by the universal sine kernel in the large $L$ limit, still sees the steady linear rise $\mathcal{O}(t)$ at intermediate time scales. This fact reaffirms the expectation that the decay in the disconnected piece, the Fourier transformed one-point functions, is model dependent. However, the ramp in the connected 2-point function is a universal feature of quantum chaotic systems.

In addition to a hidden dip, another difference with the GUE result is the lack of an oscillating decay in the LUE at infinite temperature. In the GUE, the Bessel function decay at $\beta=0$ gives a true dip time $\op(1)$. The envelope of this decay was what we considered as the decay to a dip given that a finite $\beta$ smoothed out the oscillations. 

\subsection{Two-point form factor at finite temperature}
Now let us consider the two point form factor at finite temperature. For small $\beta$, one may effectively insert the one point distribution in the integration formula. We walk through the computation in some detail as it will mimic the calculation of the supersymmetric SYK form factor in Sec.~\ref{sec:SYKch}. To be concrete, we write
\begin{align}
&\CR_2(t,\beta ) = \int D\lambda \sum_{i,j} e^{i(\lambda_i - \lambda_j)t} e^{-\beta (\lambda_i + \lambda_j)} \nn
&\quad = L \int d\lambda\, \rho(\lambda) e^{-2\beta \lambda} + L(L-1) \int d\lambda_1 d\lambda_2 \, \rho^{(2)} (\lambda_1,\lambda_2) e^{i(\lambda_1 - \lambda_2)t} e^{-\beta (\lambda_1 + \lambda_2)} \nn
&\quad = L \int d\lambda\, \rho(\lambda ) e^{-2\beta \lambda} + \int d\lambda_1 d\lambda_2\,  \Big(K(\lambda_1, \lambda_1) K(\lambda_2, \lambda_2) - K^2 (\lambda_1, \lambda_2) \Big) e^{i(\lambda_1 - \lambda_2) t} e^{-\beta (\lambda_1 + \lambda_2)} \nn
&\quad = L r_1 (2i\beta ) + L^2 r_1 (t+i\beta ) r_1(-t+i\beta ) - \int d\lambda_1 d\lambda_2 K^2 (\lambda_1, \lambda_2) e^{i(\lambda_1-\lambda_2)t} e^{-\beta (\lambda_1+\lambda_2)} \,,
\label{eq:R2bcomp}
\end{align}
simply integrating the kernels as specified above. For the final integral, we make the change of variables
\begin{equation}
u_1 = \frac{1}{2} (\lambda_1+\lambda_2) \and u_2=\lambda_1 - \lambda_2\,,
\end{equation}
which allows us to compute
\begin{align}
& \int d\lambda_1 d\lambda_2 \, K^2 (\lambda_1, \lambda_2) e^{i(\lambda_1-\lambda_2)t} e^{-\beta (\lambda_1+\lambda_2)} = \int d u_1 d u_2 \, \left( \frac{\sin (L\pi u_2)}{\pi u_2} \right)^2 e^{i u_2 t - 2\beta u_1} \nn
& \approx \int d u_1\, e^{-2\beta u_1} \rho(u_1) \int d u_2 \left( \frac{\sin (L\pi u_2) }{\pi u_2} \right)^2 e^{i u_2t} = L r_1 (2i\beta) r_2(t)\,,
\end{align}
where we regulate the unbounded integral with the insertion of $\rho(u_1)$. The 2-point spectral form factor at finite temperature is
\begin{equation}
\CR_2(t,\beta ) = L r_1(2i\beta) + L^2 r_1(t+i\beta) r_1(-t+i\beta) - Lr_1(2i\beta) r_2(t)\,.
\label{eq:R2beta}
\end{equation}

\begin{figure}
\centering
\includegraphics[width=0.95\linewidth]{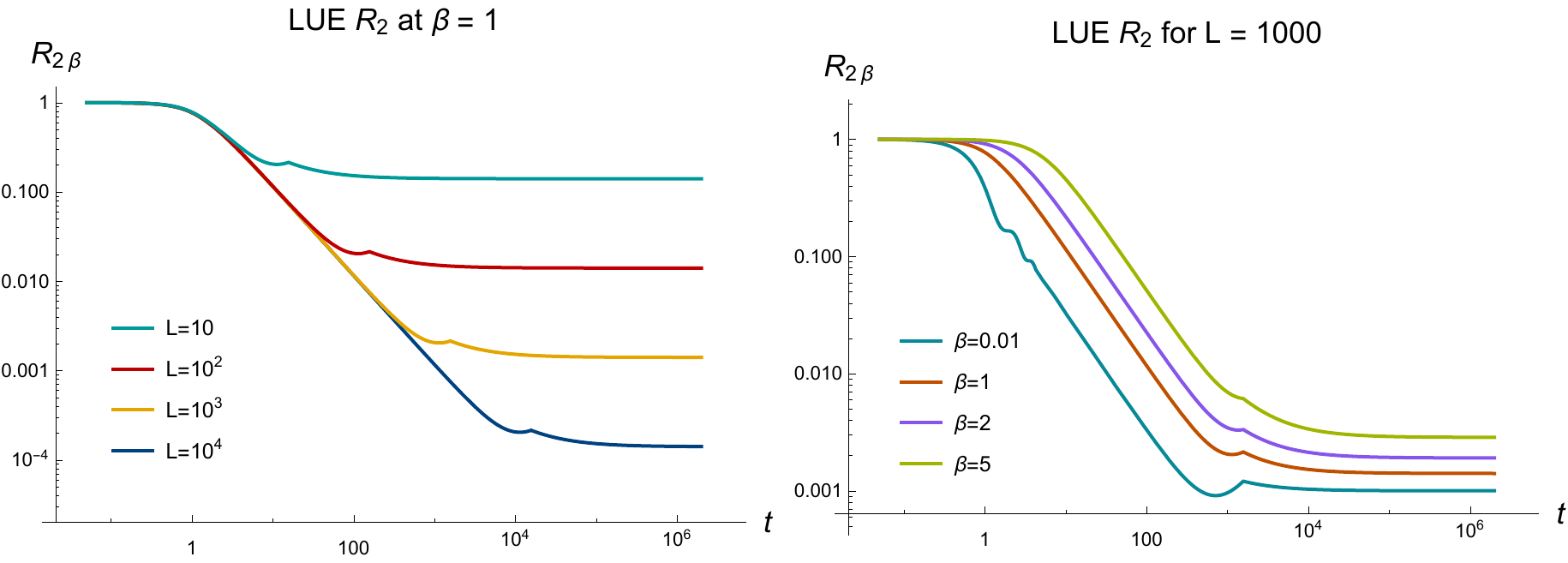}
\caption{The $2$-point spectral form factor for LUE at finite temperature, as given in Eq.~\eqref{eq:R2beta}, plotted for different values of $L$ and at different temperatures, normalized by the initial value. The plateau value depends on both $L$ and $\beta$, while the plateau time is just $L$ dependent.
}
\label{fig:LUER2beta}
\end{figure}

\noindent We plot the analytic result in Fig.~\ref{fig:LUER2beta} and observe that at finite temperature there is still no clear dip time in LUE, unlike for the GUE, and that the plateau time $t_p$ does not depend on $\beta$. For the LUE, we define $h_1(\beta) \equiv r_1(2i\beta )$, a purely real function of the inverse temperature, with the plateau value 
\begin{equation}
\CR_2 (t_p, \beta) = h_1 (2\beta) L\,.
\end{equation}
At small but finite $\beta$ we have
\begin{equation}
h_1 (2\beta) = 1 - 2\beta + 4\beta^2 + \op(\beta^3)\,,
\end{equation}
compared to the GUE result $1 + 2 \beta^2+\op(\beta^4)$ \cite{ChaosRMT}, one can see that the LUE plateau value is smaller than GUE, which is also observed in numerics.

\subsection{Four-point form factor at infinite temperature}
As an example of a higher point form factor, we compute the 4-point $\mathcal{R}_4$ at infinite temperature. By definition we have
\begin{equation}
\CR_4(t) \equiv \big\langle Z(t)Z(t)Z(t)^* Z(t)^*\big\rangle_{\rm LUE} = \int D\lambda\, \sum_{i,j,k,\ell} e^{i(\lambda_i+\lambda_j-\lambda_k-\lambda_\ell)t}\,.
\end{equation}
%\begin{itemize}
%\item $a=b=c=d=e=f$: Contribute $L$.
%\item $a=b$: Contribute $L(L-1)(L-2)\int{D\lambda}{{e}^{i(2{{\lambda }_{1}}-{{\lambda }_{2}}-{{\lambda }_{3}})t}}$.
%\item $c=d$: Contribute $L(L-1)(L-2)\int{D\lambda}{{e}^{i({{\lambda }_{1}}+{{\lambda }_{2}}-2{{\lambda }_{3}})t}}$.
%\item $a=c$ or $a=d$ or $b=c$ or $b=d$: Contribute $4L(L-1)(L-2)\int{D\lambda}{{e}^{i({{\lambda }_{1}}-{{\lambda }_{2}})t}}$.
%\item $b=c=d$ or $a=c=d$ or $a=b=d$ or $a=b=c$: Contribute $4L(L-1)\int{D\lambda}{{e}^{i({{\lambda }_{1}}-{{\lambda }_{2}})t}}$.
%\item $a=b$ and $c=d$:  Contribute $L(L-1)\int{D\lambda}{{e}^{i(2{{\lambda }_{1}}-2{{\lambda }_{2}})t}}$.
%\item $a=c$ and $b=d$, or $a=d$ and $b=c$: Contribute $2L(L-1)$.
%\item All indexes that are not equal: $L(L-1)(L-2)(L-3)\int{D\lambda}{{e}^{i({{\lambda }_{1}}+{{\lambda }_{2}}-{{\lambda }_{3}}-{{\lambda }_{4}})t}}$.
%\end{itemize}
To evaluate the expression we must consider all possible ways in which the eigenvalues can collide in the sum, \ie all equal, $\lambda_i = \lambda_j$, $\lambda_k=\lambda_\ell$, etc, and treat them separately. Making use of the 2-point form factors we derived above, and computing the 3 and 4-point function contributions by expanding the determinant and integrating products of kernels as Eq.~\eqref{eq:int}, we obtain
\begin{align}
\CR_4(t) 
&= L^4 |r_1(t)|^4 - 2L^3 {\rm Re}(r_1^2(t)) r_2(t) r_3(2t) - 4L^3 |r_1(t)|^2 r_2(t) + 2L^3 {\rm Re} (r_1(2t) r_1^{*2}(t))\nn
&\qquad +4L^3 |r_1(t)|^2 + 2L^2 r_2^2(t) + L^2 r_2^2(t) r_3^2(2t) + 8L^2 {\rm Re} (r_1(t)) r_2(t) r_3(t) \nn
&\qquad - 2L^2 {\rm Re} (r_1(2t)) r_3(2t) r_2(t) - 4L^2 {\rm Re} (r_1^* (t)) r_3(t) r_2(2t) + L^2 |r_1(2t)|^2 \nn
&\qquad - 4L^2 |r_1(t)|^2 - 4L^2 r_2(t) + 2L^2 - 7L r_2 (2t) + 4Lr_2(3t) + 4Lr_2(t) - L\,.
\end{align}
In the large $L$ limit, some of the terms above are subdominant or suppressed in $L$ at all times, allowing us to simplify the expression as
\begin{equation}
\CR_4(t) \approx L^4 |r_1(t)|^4 + 2L^2 r_2^2(t) - 4L^2 r_2(t) + 2L^2 - 7L r_2(2t) + 4L r_2(3t) + 4Lr_2(t) - L\,,
\end{equation}
similar to the result we derived for the GUE \cite{ChaosRMT}. At times much earlier than the plateau time, we have
\begin{equation}
\CR_4 \approx L^4 |r_1(t)|^4 + \frac{t(t-2\pi \rho (u))}{2\pi^2 \rho (u)^2} \sim \frac{L^4}{\pi^2 t^2} + \frac{t(t-2\pi \rho (u))}{2\pi^2 \rho (u)^2}\,.
\end{equation}
Again, we find a slow decay of $\mathcal{O}(1/t^2)$ and thus no visible dip at large $L$. The plateau time is still $2\pi \alpha_L$, with a plateau value $\CR_4(t_p) = 2L^2-L\sim 2L^2$. 

\section{Chaos and Wishart matrices}
\label{sec:LUEch}
We want to study the chaotic nature of time-evolution by LUE Hamiltonians. Consider the ensemble of unitary time-evolutions generated by LUE random matrices
\begin{equation}
\CE_t = \big\lbrace e^{-iHt}\,, ~{\rm with}~ H\in {\rm LUE}\big\rbrace\,.
\end{equation}
We want to understand how random LUE time-evolution is by asking when the ensemble forms a $k$-design. Computing the frame potential for the ensemble quantifies a distance to Haar-randomness. We also compute correlation functions of operators evolved by the LUE to look at early-time chaos in the chaotic decay of $2k$-point functions. %We will need to make use of the spectral form factors derived in Sec.~\ref{sec:SFF}. 

\subsection{QI overview}
\label{sec:QIrev}
Before discussing the frame potential and measures of chaos for the random matrix ensemble, we will briefly overview the quantum information theoretic concepts and tools we use, namely the notion of a unitary $k$-design and the frame potential. For a more in-depth review of these in the context of information scrambling in chaotic systems, see \cite{ChaosDesign,ChaosRMT}. 

For a finite dimensional quantum mechanical system, with Hilbert space $\CH$ of dimension $L$, the unitary group $U(L)$ can be equipped with the Haar measure, the unique left/right invariant measure on $U(L)$. Given some ensemble of unitary operators $\CE $, we say that the ensemble forms a unitary $k$-design if it reproduces the first $k$-moments of Haar
\begin{equation}
\int_{\rm Haar} dU\, (U^{\otimes k})^{\dagger} (\cdot) U^{\otimes k} = \int_{V\in \CE} dV\, (V^{\otimes k})^{\dagger} (\cdot) V^{\otimes k}\,,
\end{equation}
for any operator. More intuitively, we should think of this as capturing how random the ensemble is, in that the ensemble is sufficiently spread out over the unitary group to reproduce its statistics. A precise measure of Haar-randomness is the frame potential \cite{Scott08}, which measures the 2-norm distance between the $k$-th moments of an ensemble $\CE$ and Haar. The $k$-th frame potential is defined with respect to an ensemble $\CE$ as
\begin{equation}
\CF^{(k)}_\CE \equiv \int_{U,V \in \CE} dU dV\, \big| \Tr (U^\dagger V) \big|^{2k}\,.
\end{equation}
The frame potential for any ensemble $\CE$ is lower bounded by the Haar value
\begin{equation}
\CF^{(k)}_\CE \geq \CF^{(k)}_{\rm Haar}\,,
\end{equation}
with equality iff $\CE$ forms a $k$-design. The $k$-frame potential for the Haar ensemble is simply $\CF^{(k)}_{\rm Haar} = k!$ for $k\leq L$. 

The frame potential appeared in the context of information scrambling and black holes as the average of all out-of-time ordered correlators \cite{ChaosDesign}
\begin{equation}
\frac{1}{L^{4k}} \sum_{A\text{'s}, B\text{'s}} \Big| \big\langle A_1 B_1(t)\ldots A_k B_k (t) \big\rangle_\CE \Big|^{2k} = \frac{1}{L^{2(k+1)}} \CF^{(k)}_\CE\,,
\end{equation}
where ``$B(t)$'' $= U B U^\dagger$ and $U\in \CE$, averaged over any ensemble of unitaries $\CE$, with each $A_i$ and $B_i$ summed over all Pauli operators. This makes precise an approach to randomness, where the chaotic decay of correlators at late times means the frame potential becomes small and the ensemble forms a $k$-design. 

\subsection{Frame potentials}
\label{sec:FP}
\subsection*{First frame potential at $\beta=0$}
We start by computing the first frame potential at infinite temperature $\CF^{(k)}_\CE$ for the ensemble of LUE time-evolutions. Following \cite{ChaosRMT}, we have
\begin{equation}
\CF^{(k)}_{\rm LUE} = \int dH_1 dH_2 \, e^{-\frac{L}{2} \Tr H^2_1}e^{-\frac{L}{2} \Tr H^2_2} \big| \Tr \big( e^{i H_1 t} e^{-i H_2 t}\big) |^2\,.
\end{equation}
Using the unitary invariance of the ensemble and integrating using the second moment of the Haar ensemble, we find
\begin{equation}
\CF^{(k)}_{\rm LUE} = \frac{1}{L^2-1} \big( \CR_2^2 + L^2 - 2\CR_2\big)\,,
\end{equation}
with the same dependence on the form factors as in the GUE case. 

\begin{figure}
\centering
\includegraphics[width=0.95\linewidth]{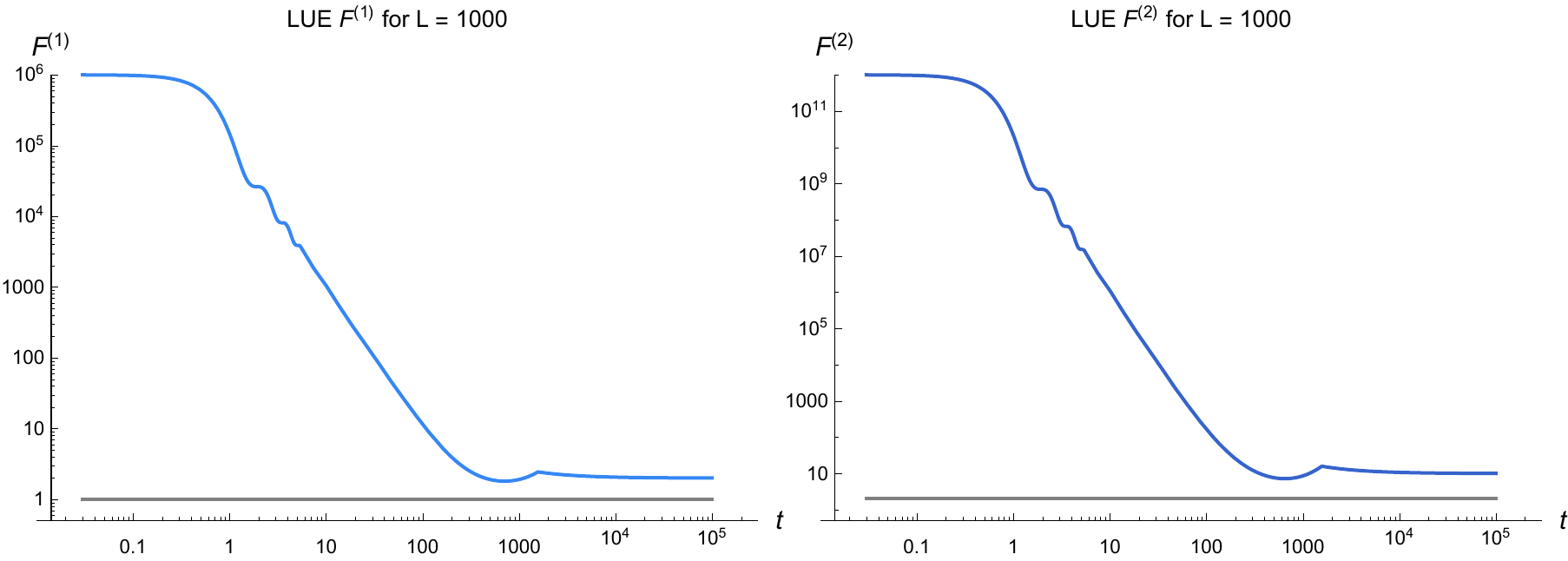}
\caption{We show the first and second frame potentials for the LUE at infinite temperature at $L=1000$. The slow decay means we do not form a $k$-design at the dip time. For comparison, the Haar value is plotted in grey.
}
\label{fig:LUEFP}
\end{figure}

In Fig.~\ref{fig:LUEFP} we plot our analytic form of the first frame potential of the LUE at infinite temperature. We can see that there are significant differences between the supersymmetric and non-supersymmetric cases. The slow decay of the LUE means there the ensemble does not form a $k$-design at the dip. At late times, after the plateau time, we find the frame potential approaches a value of 2. 

\subsection*{First frame potential at finite $\beta$}
We can also generalize the frame potential to finite temperature by averaging over all thermal $2k$-point functions with operators spaced equidistant on the thermal circle (\ie inserting $\rho^{1/2k}$ between operators in the $2k$-OTOC). Averaging over operators, we find \cite{ChaosDesign}
\begin{equation}
\CF^{(k)}_{\CE_\beta} = \int_{\CE} dH_1 dH_2\, \frac{\big| \Tr \big(e^{-(\beta /2k-it) H_1} e^{-(\beta /2k+it)H_2}\big) \big|^{2k}}{\Tr (e^{-\beta H_1}) \Tr (e^{-\beta H_2} ) / L^2}\,,
\end{equation}
with the normalization that gives the standard frame potential as $\beta\ra 0$. For the LUE, we compute the finite temperature frame potential just as above, Haar integrating to find
\begin{equation}
\CF^{(1)}_{\rm LUE}(t,\beta) = \frac{1}{L^2-1} \Big( \widetilde{\CR}_2^2(\beta/2) + L^2 - 2\widetilde{\CR}_2 (\beta /2) \Big)\,,
\end{equation}
where we define a slightly more conveniently normalized form factor
\begin{equation}
\widetilde{\CR}_2 (t, \beta)=\int D\lambda \, \frac{\sum_{ij} e^{it(\lambda_i - \lambda_j)} e^{-\beta (\lambda_i+\lambda_j)} }{ \sum_i e^{-2\beta \lambda_i}/L }\,.
\end{equation}
As it is more analytically tractable, we opt to separately average the numerator and denominator (the `quenched' version), and checked numerically that the results are in good agreement. We see that at early times, near $t=0$, we have the $\beta$-dependent value
\begin{equation}
\CF^{(1)}_{\rm LUE}\approx L^2 \frac{h_1(\beta /2)^4}{h_1(\beta )^2}\,,
\end{equation}
while at late times, after the plateau time, we have $\CF^{(1)}_{\rm LUE} (t_p,\beta )=2$. 

\subsection*{Second frame potential at $\beta=0$}
The second frame potential for the LUE at infinite temperature is expressed in terms of the spectral form factors as  \cite{ChaosRMT}
\begin{align}
\CF_{\rm LUE}^{(2)} &= \frac{1}{(L^2-9)(L^2-4)(L^2-1)L^2}\bigg( \left(L^4-8 L^2+6\right) \CR_4^2 + 4 L^2 \left(L^2-9\right) \CR_4\nn
&\quad + 4 \left(L^6-9 L^4+4 L^2+24\right) \CR_2^2-8 L^2 \left(L^4-11 L^2+18\right) \CR_2 -4 L^2 \left(L^2-9\right) \CR_{4,2} \nn
&\quad  +\left(L^4-8 L^2+6\right) \CR_{4,2}^2 + 2 \left(L^4-7 L^2+12\right) \CR_{4,1}^2 -8 \left(L^4-8 L^2+6\right) \CR_2 \CR_4\nn
&\quad -4 L \left(L^2-4\right) \CR_4 \CR_{4,1} +16 L \left(L^2-4\right) \CR_2 \CR_{4,1} -8 \left(L^2+6\right) \CR_2 \CR_{4,2}\nn
&\quad +2 \left(L^2+6\right) \CR_4 \CR_{4,2}-4 L \left(L^2-4\right) \CR_{4,1} \CR_{4,2}+2 L^4 \left(L^4-12 L^2+27\right) \bigg)\,,
\end{align}
where we have defined
\begin{equation}
\CR_{4,1}(t) \equiv \int D\lambda \sum_{i,j,k=1}^{L} e^{i(\lambda_i+\lambda_j - 2\lambda_k )t}\,, \qquad \CR_{4,2} (t) \equiv \int D\lambda \sum_{i,j=1}^{L} e^{2i(\lambda_i- \lambda_j)t}\,.
\end{equation}
The 4-point form factor with two coincident eigenvalues, $\CR_{4,2} (t)$, is simply $\CR_{2} (2t)$. The 3-point form factor $\CR_{4,1} (t)$ for the LUE can be computed just as in Sec.~\ref{sec:SFF}, where we find
\begin{align}
\CR_{4,1} (t) &= L^3 {\rm Re} \big(r_1(2t) r_1^{*2}(t) \big) - L^2 {\rm Re} (r_1(2t)) r_3(2t) r_2(t) - 2L^2 {\rm Re} (r_1^*(t)) r_3(t) r_2(2t)\nn &\quad + L^2 |r_1(2t)|^2 + 2L^2 |r_1(t)|^2 + 2Lr_2(3t) - L r_2(2t) - 2Lr_2(t) + L\,.
\end{align}
We plot the second frame potential for LUE alongside the first frame potential in Fig.~\ref{fig:LUEFP}. The second frame potential has an initial value of $L^4$ and late-time value of $10$, just as for the GUE. But again the difference arises at intermediate time scales, where the LUE fails to form a $k$-design. 

\subsection{Correlation functions}
\label{sec:otocs}
As we discussed before, the recent interest in quantum chaos has involved extensive discussion of out-of-time order correlation functions (OTOCs). Namely, the following 4-point functions of pairs of operators in thermal states
\begin{equation}
\vev{A B(t) A B(t)}_\beta \where B(t) = e^{-iHt} B e^{iHt}\,.
\end{equation}
We consider OTOCs with operators evolved by LUE Hamiltonians and averaged over the random matrix ensemble. In \cite{ChaosRMT}, we studied $2k$-OTOCs and related them to spectral quantities, both by averaging over the operators in the correlation function or over an ensemble of Hamiltonians. In that work, we averaged $2k$-OTOCs over the GUE and related the correlators to spectral quantities using the unitary invariance of the measure. As the LUE is similarly invariant, the relation between correlation functions averaged over the random matrix ensemble and the form factors will be the same as thus parts of the discussion here will closely follow \cite{ChaosRMT}; the differentiating aspects of LUE time-evolution thus lie in the spectral form factors themselves.  

First we look at the 2-point function and integrate over Hamiltonians drawn from the LUE, using the unitary invariance of the measure and Haar integrating in the eigenvalue basis
\begin{equation}
\vev{A B(t)}_{\rm LUE} = \int dH \vev{A B(t)} = \frac{\CR_2(t)-1}{L^2-1} \vev{AB}_c + \vev{A}\vev{B}\,,
\end{equation}
where $\vev{AB}_c $ denotes the connected correlator. For non-identity Paulis, the expression is nonzero for $B = A^\dagger$, and thus
\begin{equation}
\text{LUE average}:\quad \vev{A A^\dagger (t)}_{\rm LUE} \approx \frac{\CR_2(t)}{L^2}\,,
\label{eq:LUE2pt}
\end{equation}
for $\CR_2(t)\gg 1$. We note that, just as is the case for GUE, if we instead average the same 2-point function over all operators $A$, we arrive at the same expression
\begin{equation}
\text{Operator average}:\quad \int dA\, \vev{A A^\dagger (t)} = \frac{\CR_2(t)}{L^2} \,,
\end{equation}
which is true regardless of the Hamiltonian. The fact that the LUE averaged 2-point function equals the operator averaged correlator means that LUE does not care about the size or locality of the operator $A$, given that we made no assumptions about about $A$ in computing Eq.~\eqref{eq:LUE2pt}, and thus is blind to phenomena relevant for early-time chaos such as operator growth. 

We next compute the 4-point OTOC averaged over the LUE, using the fourth moment of Haar and looking at the leading order behavior
\begin{equation}
\vev{A B(t) A B(t)}_{\rm LUE} = \int dH\, \vev{A B(t) A B(t)} \approx \frac{\CR_4(t)}{L^4}\,,
\end{equation}
for non-identity Pauli operators $A$ and $B$. Note that the OTOCs of the form $\vev{A B(t) C D(t)}$ are all almost zero unless $ABCD = I$. 

We can now comment on the time scales that LUE describes as seen from the averaged correlation functions. The time scale of 2-point function decay corresponds to the time scales for which the system thermalizes. Using the early time piece of the 2-point form factor we derived in Sec.~\ref{sec:SFF}, where the contribution from the 1-point function gives the decay
\begin{equation}
\vev{A A^\dagger (t)}_{\rm LUE} \approx J_0^2 (2t) + J_1^2 (2t) \sim \frac{1}{\pi t} \,,
\end{equation}
contrasted to the $1/t^3$ decay for GUE. Similarly, we can comment on scrambling in the LUE by looking at the early time decay of the LUE averaged 4-point OTOCs. The early time behavior of the 4-point form factor means the OTOC decays like
\begin{equation}
\vev{A B(t) A B(t)}_{\rm LUE} \approx \big( J_0^2 (2t) + J_1^2 (2t) \big)^2 \sim \frac{1}{\pi^2 t^2}\,.
\end{equation}

The characteristic time-scale for decay of LUE 2-point functions is $t_2 \sim \op(1)$, or for systems at finite temperature $\op (\beta)$. The time-scale for 4-point function decay is also order 1, but faster than the decay of 2-point functions $t_4 \sim t_2/2$. Although the decay is slower than for GUE, unsurprisingly, the conclusion about the LUE's perception of early-time chaos is the same: the LUE 4-point OTOCs decay faster than the LUE 2-point functions, which means the random matrix ensemble fails to describe scrambling at early times. 

\subsection{Complexity}
Lastly, we briefly comment on the complexity growth under time-evolution of LUE Hamiltonians. Here we simply discuss the results; details and definitions of ensemble complexity and its relation to the frame potential are given in \cite{ChaosDesign,ChaosRMT}. The gate complexity of an ensemble $\CE$, \ie the number of gates needed to generate $\CE$, is lower bounded by the frame potential as
\begin{equation}
\CC(t) \geq \frac{2kn - \log \CF^{(k)}_\CE(t)}{2 \log n}\,.
\end{equation}
At early times before the dip time $t\ll t_d$, the dominant contribution to the $k$-th frame potential is $\CF^{(k)}_\CE\simeq \CR_{2k}^2(t) / L^{2k} $ \cite{ChaosRMT}. For $k\ll L$, the 2$k$-th form factor goes as $\CR_{2k} \sim r_1^{2k}$, the function defined in Eq.~\eqref{eq:rfuncs} in terms of Bessel functions. The decay $r_1^2 \sim 1/t$, gives a lower bound on the growth of the circuit complexity
\begin{equation}
\CC(t) \geq \op \left( \frac{k \log t}{\log n}\right)\,,
\end{equation} 
where the slower decay for LUE still gives the same logarithmic lower bound as GUE. Interestingly, in GUE the 1-point function contribution to the form factor at early times is an oscillating Bessel function decay $J^2_0(2t)/t^2$, which formally gives a dip time $\op(1)$. As these oscillations are not present in the LUE, we can bound the complexity up to the dip time even at infinite temperature. But for large $k$, we recover the quadratic growth of complexity: $\CC \geq t^2 /\log n$, hinting again at the unphysical nature of LUE evolution at early times.

\section{Chaos in supersymmetric SYK}
\label{sec:SYKch}
The supersymmetric SYK model admits a classification by Wishart-Laguerre random matrix ensembles and has a density of states which closely follows a Mar{\v c}enko-Pastur distribution \cite{SUSY_RMT}. Having discussed the properties of LUE random matrices, we turn to the supersymmetric SYK model and check that the form factor acts similarly. From the frame potential, we then discuss the Haar-randomness of the model's time evolution.

Assuming that the spectral statistics of the theory are Gaussian, as both SYK and the Wishart matrices are, allows us to use the sine kernel to compute the spectral $n$-point functions. We note that if the statistics are GUE/GOE/GSE, the sine kernel is slightly modified and the ramp function differs as we approach $t\sim L$, but the universal growth of the ramp is still present. Knowing that the supersymmetric SYK model has Gaussian spectral statistics \cite{SUSY_RMT}, we can compute the finite temperature form factor for the theory just as in Eq.~\eqref{eq:R2bcomp}, and find
\begin{align}
\CR_2(t,\beta) &= \big\langle Z(\beta+it) Z(\beta-it) \big\rangle = \int D\lambda\, \sum_{i,j} e^{i(\lambda_i-\lambda_j)t} e^{-\beta(\lambda_i+\lambda_j)}\nn
& \approx L \int dE\, \rho(E) e^{-2\beta E} + \big| \vev{Z(\beta+it)} \big|^2 - L \int dE\, e^{-2\beta E} \rho(E) r_2(t)\,,
\end{align}
where $r_2(t)$ is the ramp function from the LUE and we define $E = \frac{1}{2}(\lambda_1+\lambda_2)$. Continuing, we find the finite temperature form factor
\begin{equation}
\CR_2(t,\beta) \approx | \vev{Z(\beta+it)} |^2 + Z(2\beta) \big(1-r_2(t)\big)\,.
\end{equation}
As a sanity check, the late-time value $Z(2\beta)$ here matches the infinite-time average of the spectral form factor. As we discussed in Sec.~\ref{sec:setup}, the 1-loop partition function from the super-Schwarzian theory is
\begin{equation}
Z^\text{sSch}_\text{1-loop}(\beta) \sim \frac{1}{\sqrt{\beta \CJ}} e^{Ns_0 + cN/2\beta}\,,
\end{equation}
where $s_0$ is the ground-state entropy density and $c$ is the specific heat. At early times, the form factor is dominated by its disconnected component, decaying as $1/t$
\begin{equation}
{\rm Early}:\quad \CR_2(t,\beta) \sim \frac{e^{2N s_0}}{\CJ t}
\end{equation}
for times greater than $t\sim \sqrt{N} = \log L/2$, but shorter than $t\sim \sqrt{L}$. Computing the connected form factor, we find
\begin{align}
\CR_2^c(t,\beta) &\equiv \big\langle Z(\beta +it)Z(\beta -it) \big\rangle - |\vev{ Z(\beta +it)}|^2 \nn
& =Z(2\beta) \big(1- r_2(t)\big) = \frac{1}{\sqrt{2\beta \CJ}} e^{Ns_0+cN/4\beta} \big(1-r_2(t)\big)\,.
\end{align}
Equating the $1/t$ decay with the ramp gives a dip time $t_d \sim e^{Ns_0}$, the same order as the plateau time $t_p$. Even in light of the exactness of the super-Schwarzian theory, we should be cautious in extrapolating to very late times. It is possible that in the large $N$ theory the slope is not well-described by the effective theory at late times and, in turn, decays faster at an intermediate time scale. 

\begin{figure}
\centering
\includegraphics[width=0.95\linewidth]{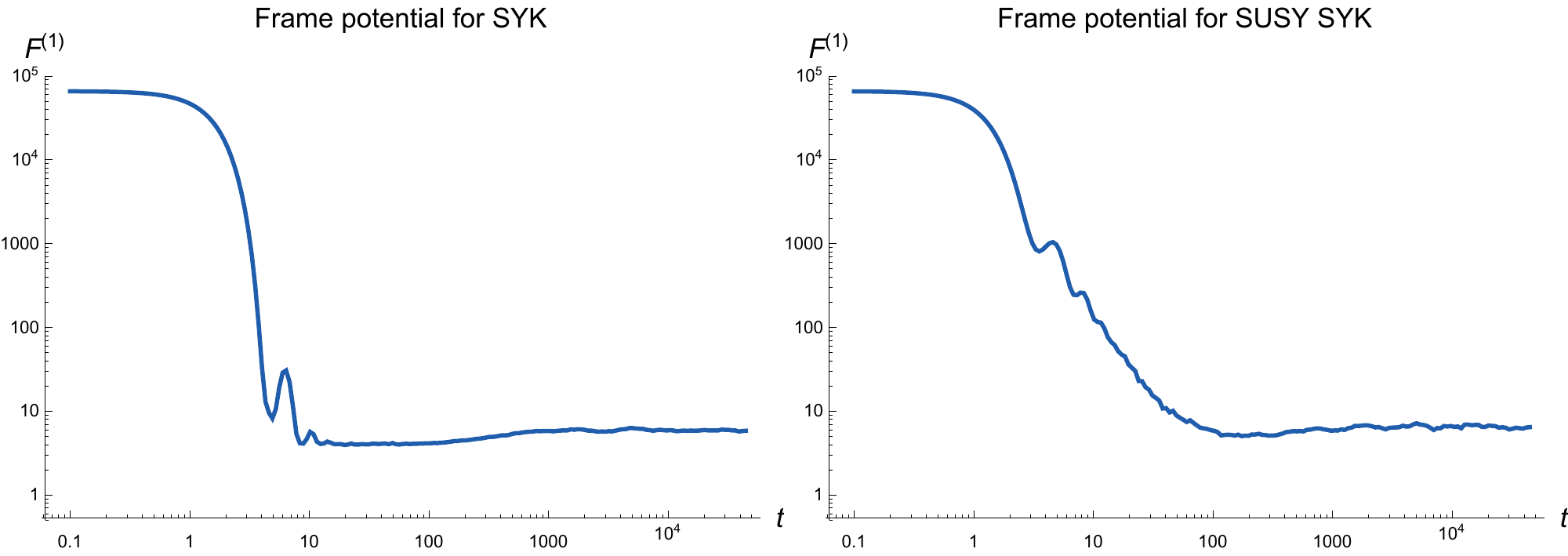}
\caption{Numerics for the first frame potential of SYK and supersymmetric SYK at $\beta=0$ for $N=16$ Majoranas and 200 samples. The decay and dip of SYK indicates faster scrambling and an approximate $k$-design behavior not as readily apparent in the supersymmetric model.
}
\label{fig:SYKFP}
\end{figure}

Lastly, to get a hint at the nature of scrambling and an approach to randomness in SYK and its supersymmetric extension, we numerically plot the first frame potential for each in Fig.~\ref{fig:SYKFP} at infinite temperature and for $N=16$ Majoranas. The faster decay and dip that appears for SYK means the frame potential decays quickly, forming an approximate $k$-design at the dip time. Although the dip value of the SYK frame potential for $N=16$ is larger than the Haar value, we checked that as we increase $N$ the dip value decreases and expect that SYK forms an approximate $k$-design in the large $N$ limit. The frame potential for the supersymmetric model exhibits a much more gradual approach to its minimal value which is larger than in SYK, indicating less effective information scrambling and a greater distance of the ensemble to forming a $k$-design. It would be interesting to see, either numerically or analytically, if these behaviors persist at large $N$.  Both theories, like their random matrix counterparts, become less random and increase after the dip, deviating further from an approximate design, which suggests that $k$-invariance \cite{ChaosRMT} might provide a better insight in how information scrambles in SYK models.

There are a few comments worth making relating the discussion here with the behavior of the form factor in similar models.\footnote{We thank an anonymous JHEP referee for raising these points.}
In the complex SYK model, the spectral form factor appears to have a $1/t^4$ power-law decay at early times \cite{Davison16,Sonner17}, in contrast to the Majorana and SUSY SYK models. As we discussed, the respective power-law decays in these models arise from the Schwarzian and super-Schwarzian modes governing the low-energy physics, and persist for a long time as a result of the 1-loop exactness of the effective actions. In the complex SYK model, where we have a conserved $U(1)$, there is an additional contribution to the effective action from the phase fluctuations of the reparametrization mode, as was discussed in \cite{Davison16}. Combined with the contribution from the Schwarzian mode, the partition function has a $Z(\beta)\sim 1/(\beta \CJ)^2$ dependence. Continuing to real-time, the early-time contribution to the 2-point form factor gives a power-law decay $\CR_2(t) \sim |Z(\beta,t)|^2 \sim 1/t^4$. As the low-energy description is likely also 1-loop exact, one expects this behavior to persist for a long time. It is further interesting to note that while the power-law indicates a more rapid onset of late-time chaos as seen by the frame potential, the additional $U(1)$-mode does not contribute to the Lyapunov exponent of the theory \cite{Bulycheva17}. Thus, like Majorana SYK and SUSY SYK models, the complex SYK model is maximally chaotic at early times, but in the above sense scrambles quicker.

We should also comment on the behavior of spectral quantities more generally in chaotic systems with gravitational duals. In 2d CFTs, an analysis of the contribution from different saddles indicates a persisting $1/t^3$ decay in the form factor for holographic CFTs, and a $1/t$ decay for rational CFTs \cite{CFTlatetimes}.\footnote{Relatedly, \cite{Asplund15} discussed a distinction between entanglement scrambling in rational and holographic CFTs.} A slow decay of spectral quantities also appears in the D1-D5 theory at the orbifold point, in line with the fact that the theory does not have chaotically decaying correlation functions \cite{Perlmutter16} and appears to exhibit a logarithmic ramp \cite{D1D5chaos}, in contrast to the universal linear ramp we expect in chaotic systems. Although \cite{BHRMT16} argued for the rapid decay of spectral functions and the late-time appearance of a ramp in super Yang-Mills at strong-coupling, better analytic control of spectral quantities is needed to understand quantum chaos in holographic theories.

\section{Conclusion and outlook}
\label{sec:con}

In this paper, we considered the Wishart-Laguerre unitary ensemble in order to understand universal features of supersymmetric quantum mechanical systems. We computed the 2-point spectral form factor for the LUE and found the one-point function contribution gives a $1/t$ power law decay at early times, hiding the dip and transitioning directly into the plateau. This is relatively slow compared to the $\sim 1/t^3$ decay seen in both SYK and the GUE. The universal ramp behavior from the sine kernel can be seen in the connected LUE 2-point form factor. These results agree with the prediction from the 1-loop partition function in supersymmetric SYK. This slow decay implies the onset of a random matrix description occurs at much later times. This can best be seen from the frame potential, where we find a more gradual decay to Haar-random dynamics. Moreover, the frame potential for the LUE, unlike that of the GUE, does not reach the Haar value and does not form an approximate $k$-design. This is also what we predict and observe numerically in the supersymmetric SYK model, where the slower decay and larger dip value imply less effective information scrambling.

The supersymmetric model, while maximally chaotic, sees a slower onset of random matrix behavior---made evident by the lack of a dip in the form factor and by the slow approach to Haar-randomness in the frame potential. The apparent distinction here between early-time chaos, in terms of chaotic correlation functions, and late-time chaos, in terms of scrambling and Haar-randomness, demands a deeper understanding. 

\vspace*{24pt}
\noindent{\it Note added:} In the preparation of this draft, \cite{CL17} appeared which also considers the infinite temperature 2-point spectral form factor for Wishart matrices in a different context. Namely, they study the statistical properties of the reduced density matrix on spatial regions in quantum many-body systems. They also comment on universal features of Wishart matrices in Floquet systems. As there is a sense in which Floquet systems may be thought of as supersymmetric quantum mechanics \cite{FloqSusy}, where the Floquet unitary is built from two `supercharges', it would be interesting to explore further connections with our work.

\vspace*{12pt}
\subsection*{Acknowledgments}
We thank Beni Yoshida for helpful discussions. NHJ acknowledges support from the Simons Foundation through the ``It from Qubit'' collaboration as well as from the Institute for Quantum Information and Matter (IQIM), an NSF Physics Frontiers Center (NSF Grant PHY-1125565) with support from the Gordon and Betty Moore Foundation (GBMF-2644). JL acknowledges support from the U.S. Department of Energy, Office of Science, Office of High Energy Physics, under Award Number DE-SC0011632.

\appendix

\section{Numerics}
\label{app:num}
In this appendix we discuss numerics to fix an analytic form of the form factors for LUE and to further provide checks on the expressions we derived for the form factors and frame potentials. As we mentioned in Sec.~\ref{sec:SFF}, there was a free parameter $u$ in the expressions we derived for the $k$-point form factors. This dependence appears in the ramp function $r_2(t)$, defined in Eq.~\eqref{eq:rfuncs}, and determines both the slope of the linear ramp in $\CR^c_{2}(t)$ and the plateau time. Numerically computing the connected 2-point form factor for $L=500$, we fix $u$ by fitting the ramp between times $\sim$1 and $\sqrt{L}/2$. We know that the early time behavior of the ramp is quadratic before $t\sim 1$ and expect a loss of analytic control as we approach the plateau time. We thus linearly fit points in this intermediate regime and find $u=1.156$. We hope to derive this result more rigorously in the future. 

\begin{figure}[htb!]
\centering
\includegraphics[width=0.99\linewidth]{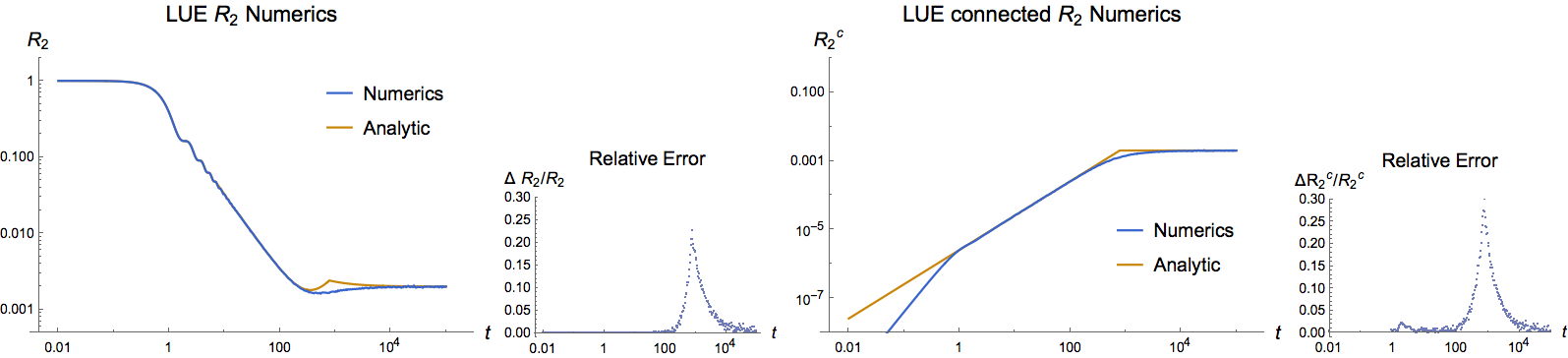}
\caption{Numerics for both the LUE 2-point form factor and its connected component, compared to the analytic expressions derived in Sec.~\ref{sec:SFF}, for $L=500$ and with 10000 samples. We find good agreement in the slope and plateau, with expected deviations around the plateau time. The very early time behavior of the connected form factor can also be understood analytically.  
}
\label{fig:R2num}
\end{figure}

We also present some numerical checks of our expressions for the LUE 2-point form factor in Fig.~\ref{fig:R2num}, where we find good agreement in the slope, ramp, and plateau.  Our results were derived for LUE at large $L$ and thus should capture the perturbative behavior. But in the transition to the plateau, nonperturbative effects \cite{AAinst} become important and our results deviate from numerics in this regime. After the plateau time, we return to contributions from the 1-point function. At very early times, before $t\sim \op(1)$, the connected component grows as $\CR_2^c(t) \sim t^2$. This quadratic growth can be derived from an impressive integral representation of the connected 2-point form factor \cite{BrezinRMT}. We have also checked our expressions of the finite temperature and higher point LUE spectral functions and found good agreement with numerics. 

\bibliographystyle{utphys}
\bibliography{chaos_susy}

\end{document}